\documentclass[10pt,aps,prb,superscriptaddress,twocolumn,showpacs,preprint]{revtex4-1}

\usepackage{graphicx}
\usepackage{dcolumn}
\usepackage{bm}
\usepackage{amsmath}
\usepackage{url}
\usepackage[T1]{fontenc}
\usepackage{xfrac}
\usepackage[dvipsnames]{xcolor}
\begin{document}

%\preprint{APS/123-QED}

\title{Element-specific soft X-ray spectroscopy, scattering and imaging studies of skyrmion-hosting compound Co$_8$Zn$_8$Mn$_4$}

\author{V. Ukleev}
\email{victor.ukleev@psi.ch}
\affiliation{RIKEN Center for Emergent Matter Science (CEMS), Wako 351-0198, Japan}
\affiliation{Laboratory for Neutron Scattering and Imaging (LNS), Paul Scherrer Institute (PSI), CH-5232 Villigen, Switzerland}
\author{Y. Yamasaki}
\affiliation{RIKEN Center for Emergent Matter Science (CEMS), Wako 351-0198, Japan}
\affiliation{Research and Services Division of Materials Data and Integrated System (MaDIS), National Institute for Materials Science (NIMS), Tsukuba, 305-0047 Japan}
\affiliation{PRESTO, Japan Science and Technology Agency (JST), Kawaguchi 332-0012, Japan}
\author{D. Morikawa}
\affiliation{RIKEN Center for Emergent Matter Science (CEMS), Wako 351-0198, Japan}
\author{K. Karube}
\affiliation{RIKEN Center for Emergent Matter Science (CEMS), Wako 351-0198, Japan}
\author{K. Shibata}
\affiliation{RIKEN Center for Emergent Matter Science (CEMS), Wako 351-0198, Japan}
\author{Y. Tokunaga}
\affiliation{Department of Advanced Materials Science, University of Tokyo, Kashiwa 277-8561, Japan}
\author{Y. Okamura}
\affiliation{Department of Applied Physics and Quantum-Phase Electronics Center (QPEC), University of Tokyo, Tokyo 113-8656, Japan}
\author{K. Amemiya}
\affiliation{Condensed Matter Research Center and Photon Factory, Institute of Materials Structure Science, High Energy Accelerator Research Organization, Tsukuba 305-0801, Japan}
\author{M. Valvidares}
\affiliation{CELLS Experiment Division, ALBA Synchrotron Light Source, Barcelona E-08290, Spain}
\author{H. Nakao}
\affiliation{Condensed Matter Research Center and Photon Factory, Institute of Materials Structure Science, High Energy Accelerator Research Organization, Tsukuba 305-0801, Japan}
\author{Y. Taguchi}
\affiliation{RIKEN Center for Emergent Matter Science (CEMS), Wako 351-0198, Japan}
\author{Y. Tokura}
\affiliation{RIKEN Center for Emergent Matter Science (CEMS), Wako 351-0198, Japan}
\affiliation{Department of Applied Physics and Quantum-Phase Electronics Center (QPEC), University of Tokyo, Tokyo 113-8656, Japan}
\author{T. Arima}
\affiliation{RIKEN Center for Emergent Matter Science (CEMS), Wako 351-0198, Japan}
\affiliation{Department of Advanced Materials Science, University of Tokyo, Kashiwa 277-8561, Japan}
%\date{\today}

\begin{abstract}
A room-temperature skyrmion-hosting compound Co$_8$Zn$_8$Mn$_4$ has been examined by means of soft X-ray absorption spectroscopy, resonant small-angle scattering and extended reference holography. An element-selective study was performed by exciting the $2p$-to-$3d$ transitions near Co and Mn $L_{2,3}$ absorption edges. By utilizing the coherence of soft X-ray beams the element-specific real-space distribution of local magnetization at different temperatures has been reconstructed using iterative phase retrieval and holography with extended reference. It was shown that the magnetic moments of Co and Mn are ferromagnetically coupled and exhibit similar magnetic patterns. Both imaging methods provide a real-space resolution of 30 nm and allowed to record a magnetic texture in the temperature range between $T\,=\,20$ K and $T\,=120\,$ K, demonstrating the elongation of the skyrmions along the principal crystallographic axes at low temperatures. Micromagnetic simulations have shown that such deformation is driven by decreasing ratio of symmetric exchange interaction to antisymmetric Dzyaloshinskii-Moriya interaction in the system and effect of the cubic anisotropy.
\end{abstract}

%\pacs{61.05.cf, 61.05.cj, 75.25.-j,75.47.Np, 42.30.Rx.}
\maketitle

\section{Introduction}

Magnetic properties of transition metals (TM) are generally determined by the $3d$ valence electrons. Resonant soft X-ray scattering at $L_{2,3}$ absorption edges of TM involves $2p$-to-$3d$ transitions, thus being an element-selective probe with possibility to distinguish magnetic signal from different elements in multicomponent magnets \cite{fink2013resonant}. Moreover, the spatial coherence of the polarized soft X-ray beams provided by modern synchrotron radiation sources and free-electron lasers give vast opportunities for the lensless imaging using coherent diffraction imaging \cite{turner2011x,flewett2012method,ukleev2018coherent}, ptychography \cite{tripathi2011dichroic,shi2016soft}, and holographic techniques \cite{eisebitt2004lensless,duckworth2011magnetic}. Both coherent resonant soft X-ray scattering (RSXS) imaging and holography allow to perform real-space imaging of the magnetization distribution in thin samples with various environments, such as in high magnetic fields or at low temperatures. Flexibility of the environment and enhanced robustness of these methods against the specimen displacements are significant advantage of the lensless techniques compared to scanning transmission magnetic X-ray microscopy (STXM) \cite{fischer2015x}, although the possibility of cryogenic STXM imaging also has been recently demonstrated \cite{simmendinger2018transmission}. Coherent diffraction allows the solution of classical crystallographic inverse problem of phase retrieval by using the iterative reconstruction algorithms applied to the resonant diffraction intensities \cite{miao1999extending}. X-ray magnetic holography is based on the utilization of the interference between magnetic scattering from the object under investigation and reference wave generated by charge scattering from the prepared source. Imaging experiments using holographic approaches can be realized in a few different ways: Fourier transform holography (FTH) is based on a reference wave scattered from one or multiple small ($30-150$\,nm) pinholes placed near the sample aperture \cite{eisebitt2004lensless}. Alternatively, holography with extended reference by autocorrelation linear differential operation (HERALDO) \cite{guizar2007holography} can be performed. In contrast to FTH, HERALDO technique implies the scattering from extended reference object, such as a narrow slit or a sharp corner, which allows to improve the contrast of the real-space image without compromising the resolution \cite{zhu2010high}. Moreover, fabrication of the extended reference is less challenging than the array of the reference pinholes \cite{zhu2010high,buttner2013automatable}. Previously, FTH and HERALDO techniques were successfully applied for imaging of the element-specific magnetic domain patterns in thin films and multilayers with perpendicular magnetic anisotropy \cite{stickler2010soft,camarero2011exploring,weder2017multi}. Both FTH and HERALDO with references milled at an oblique angle into the masks also allow the imaging at a tilted angle, which is relevant for the spintronic devices with an in-plane magnetic anisotropy, such as spin-valves \cite{duckworth2013holographic}, and magnetic nanoelements \cite{tieg2010imaging,parra2016holographic,bukin2016time}. Thus the coherent soft X-ray scattering and imaging are powerful tools to study the spin ordering in multicomponent magnetic compounds with element selectivity.
 
Recently, several bulk materials that exhibit non-trivial topological spin textures and contain two or more magnetic elements have been discovered: doped B20-type alloys \cite{adams2010skyrmion, shibata2013towards}, Co-Zn-Mn compounds with $\beta$-Mn structure \cite{tokunaga2015new}, molybdenum nitrides \cite{li2016emergence} and Heusler alloys \cite{meshcheriakova2014large,phatak2016nanoscale,nayak2017magnetic}. Competition between the magnetic interactions in non-centrosymmetric compounds results in the complex phase diagram. The interplay between exchange interaction, antisymmetric Dzyaloshinskii-Moriya interaction (DMI), and magnetocrystalline anisotropy may cause incommensurate spin phases such as helical, conical and Bloch-type skyrmion lattice states \cite{bak1980theory,rossler2006spontaneous}.
% Skyrmions lattices were previously observed in various bulk B20 compounds \cite{muhlbauer2009skyrmion,munzer2010skyrmion,yu2011near,seki2012observation,kanazawa2012possible}, Heusler alloys (both skyrmions \cite{meshcheriakova2014large}, polar magnetic semiconductors \cite{kezsmarki2015neel,fujima2017thermodynamically,bordacs2017equilibrium}. 
The typical size of a magnetic skyrmion varies in a range from a few to a few hundreds nm which makes them promising candidates for future spintronic applications such as skyrmion racetrack memory and logic devices \cite{fert2017magnetic}. The skyrmions can be manipulated by current pulses with ultra-low current densities \cite{iwasaki2013universal}, electric \cite{white2014electric,okamura2016transition} and microwave fields \cite{onose2012observation,okamura2013microwave,wang2015driving}, and temperature gradients \cite{everschor2012rotating,kong2013dynamics,mochizuki2014thermally}. In the past decade skyrmion textures have been extensively studied by means of small-angle neutron scattering (SANS) \cite{adams2011long,seki2012formation,moskvin2013complex} and Lorentz transmission electron microscopy (LTEM) \cite{yu2010real,yu2011near}. Also, several groups reported on the resonant X-ray diffraction \cite{langner2014coupled,zhang2016multidomain}, small-angle scattering \cite{yamasaki2015dynamical,okamura2017directional,okamura2017emergence,zhang2017room} studies of Bloch-type skyrmions in the chiral magnets and imaging of N\'eel-type skyrmions stabilized by interfacial DMI \cite{buttner2015dynamics,blanco2015nanoscale,Woo2016,Woo2017}. Since X-ray magnetic circular dichroism (XMCD) is sensitive to the component of the magnetization parallel to the incident X-ray beam, transmission soft X-ray imaging is a complementary method to LTEM, which is sensitive to the in-plane magnetic flux inside the sample \cite{chapman1984investigation}. Hence for the N\'eel-type skyrmions, where the curl of magnetization lies in the sample plane and produce no contrast for LTEM without tilting the sample \cite{jiang2016mobile,pollard2017observation}, soft X-ray imaging, has been successfully employed for the room-temperature N\'eel-type skyrmion-hosting thin films \cite{buttner2015dynamics,blanco2015nanoscale,Woo2016,Woo2017}, but not yet for the thin plates of polar magnets, that order magnetically at cryogenic temperatures \cite{kezsmarki2015neel,kurumaji2017neel}.

Room-temperature magnetic ordering of the chiral $\beta$-Mn-type Co-Zn-Mn alloy \cite{tokunaga2015new} makes these materials promising for applications. The $\beta$-Mn-type compound Co$_8$Zn$_8$Mn$_4$ exhibits a transition from the paramagnetic state to a helical or Bloch-type skyrmion lattice state at $T_c \approx 300$\,K with a magnetic modulation period of 125\,nm \cite{karube2016robust}, and undergoes a spin glass transition
at $T_g\approx8$\,K, probably due to freezing of Mn spins \cite{karube2018sciadv}. Frustration at the Mn site ultimately results in increment of a spin-glass transition temperature ($T_g\approx30$\,K) in the compound with higher Mn concentration Co$_7$Zn$_7$Mn$_6$. Moreover, a low-temperature frustration-induced equilibrium skyrmion phase has been recently found in the latter \cite{karube2018sciadv}. In the present work we utilized the polarization-dependent soft X-ray magnetic spectroscopy, coherent RSXS, and HERALDO techniques to perform an element-selective study of the magnetic interactions and long-range ordering in Co$_8$Zn$_8$Mn$_4$ compound. The coherence of the synchrotron radiation allowed us to successfully combine small-angle scattering in transmission geometry with coherent diffraction imaging and employ the small-angle scattering patterns for the real-space reconstruction of the local magnetization distribution via iterative phase retrieval algorithm. The coherent diffraction imaging results were compared to the real-space reconstruction results provided by HERALDO. 

\section{Experimental}

X-ray spectroscopy, scattering and imaging experiments were performed at the variable-polarization soft X-ray beamline BL-16A of the Photon Factory (KEK, Japan) \cite{amemiya2010k} and BL29 BOREAS of the ALBA synchrotron radiation laboratory (Cerdanyola del Vall\'es, Spain) \cite{barla2016design}.

\begin{figure*}
\includegraphics[width=16cm]{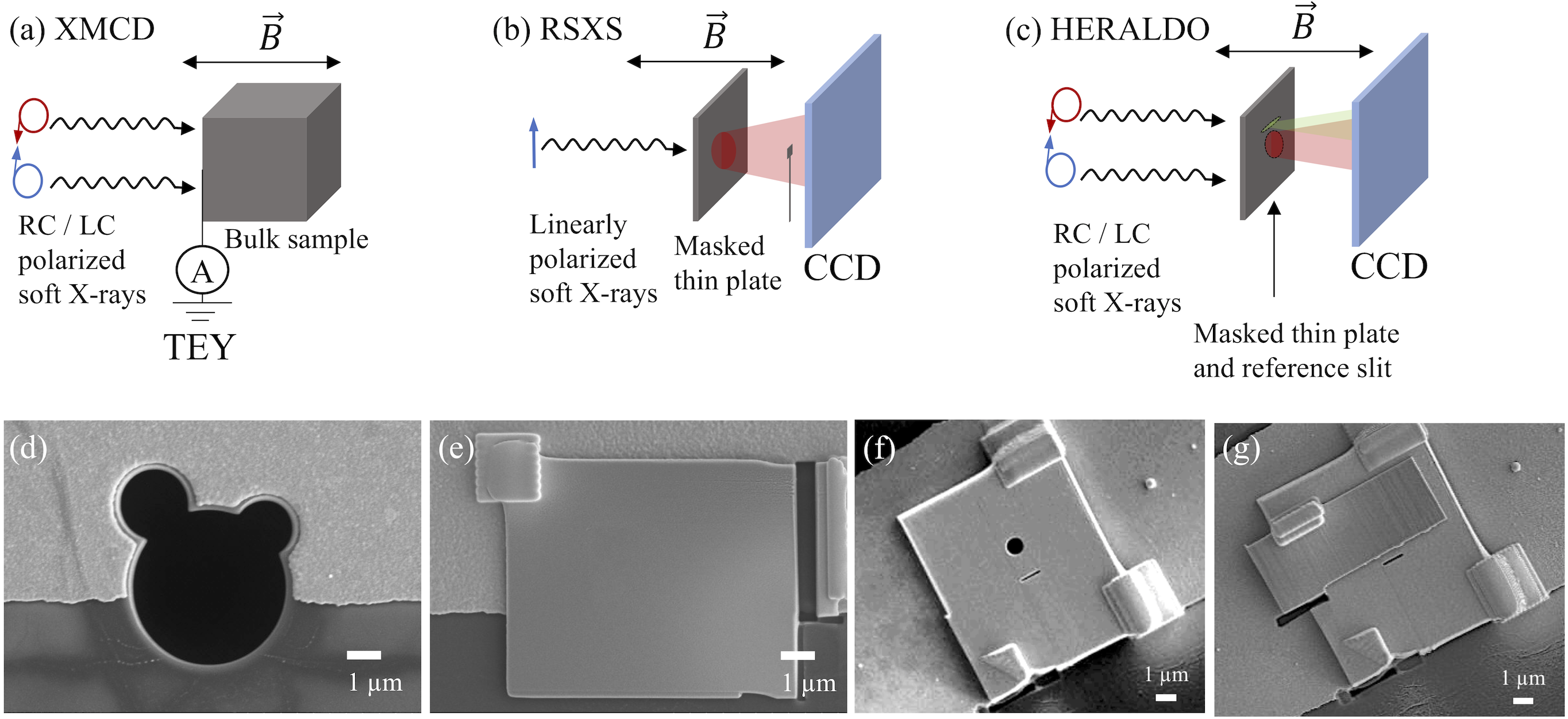}
\caption{(Color online) Sketch of the (a) XMCD, (b) RSXS and (c) HERALDO experiments. (d) SEM image of the RSXS sample aperture. (e) Thin plate of the Co$_8$Zn$_8$Mn$_4$ fixed onto the membrane. (f) SEM image of the sample aperture and the reference slit drilled in the 1-$\mu$m-thick gold plate for HERALDO experiment. (g) Co$_8$Zn$_8$Mn$_4$ thin plate fixed onto gold plate by tungsten contact.  The strong white/black contrast in the top/bottom parts of the panels (d--g) corresponds to the Au wires sputtered onto the membrane \cite{okamura2017directional}, that is irrelevant for this study.}
\label{Fig1}
\end{figure*}

Experimental geometry of soft X-ray absorption (XAS) and XMCD experiments are shown in Fig.\ref{Fig1}a. A bulk polycrystalline Co$_8$Zn$_8$Mn$_4$ specimen was obtained from an ingot grown by Bridgman method as described in Ref.\cite{karube2016robust}. The sample was polished to remove the potentially oxidized surface layer prior to the experiment. Sample was placed to the vacuum chamber with pressure of $10^{-9}$\,Torr equipped with a 5\,T superconducting magnet. XAS and XMCD signals were measured with energy resolution of 0.1\,eV using the surface-sensitive total electron yield (TEY) method near Co and Mn $L_{2,3}$ absorption edges with right and left circularly polarized (RCP and LCP) X-rays. 

RSXS and HERALDO experiments were performed in the transmission geometry (Figs. \ref{Fig1}b,c). Commercial silicon nitride membranes from Silson Ltd (Southam, UK) were processed for soft X-ray experiments. The front side of each membrane was coated with $\approx 4$\,$\mu$m-thick layer of gold to absorb the incoming X-ray beam. Further treatment of the membranes and thin plate fabrication were carried out by using focused ion beam (FIB) setup Hitachi NB5000 equipped with scanning electron microscope (SEM). Since the attenuation length for soft X-rays in Co$_8$Zn$_8$Mn$_4$ alloy is of about 100\,nm, the scattering and imaging experiments were carried out on thin plates of Co$_8$Zn$_8$Mn$_4$ with thickness of 200\,nm and 150\,nm, respectively.  Two FIB-thinned plates of a Co$_8$Zn$_8$Mn$_4$ containing (001) plane were cut from the bulk single crystal. For the RSXS experiment a thin plate was attached directly to the Si$_3$N$_4$ membrane (Figs.\,\ref{Fig1}d,e). In case of the RSXS sample the aperture with a diameter of 4.5\,$\mu$m and asymmetric shape, which provides a better convergence of the phase retrieval algorithm \cite{fienup1986phase} was drilled in the gold coating (Fig. \ref{Fig1}d). 

In the case of the HERALDO sample we fabricated the sample aperture by a different fabrication approach: a large hole with a diameter of 6\,$\mu m$ was drilled in the gold-coated membrane, and covered by 1\,$\mu$m-thick gold plate fabricated by FIB from a bulk specimen. Then, a circular sample aperture with a diameter of 700\,nm and a reference slit with a length of 1\,$\mu$m and width 40\,nm were milled in the Au plate (Fig. \ref{Fig1}f). The slit length and distance from the aperture were chosen according to the separation conditions preventing the overlapping of sample autocorrelation and sample-reference cross-correlation at the reconstruction \cite{guizar2010differentially}. 

To prevent the specimen damage by Ga$^+$ ions in the aperture and reference milling process the thin plate was fixed to the membrane by means of a tungsten contact after the mask treatment (Fig. \ref{Fig1}e). For both RSXS and HERALDO the thin plates were attached to the corresponding membranes by the single tungsten contacts (Fig. \ref{Fig1}g) to avoid the possible strain \cite{shibata2015large,okamura2017emergence}.

The RSXS setup at Photon Factory, Japan was equipped with a high-vacuum chamber with a background pressure of $10^{-8}$\,Torr \cite{yamasaki2015dynamical}. The scattered intensity was collected by an in-vacuum charge-coupled device (CCD) area X-ray detector of $2048\,\times\,2048$ pixels (Princeton Instruments, Trenton, New Jersey, USA). The RSXS endstation MARES was used at ALBA synchrotron \cite{barla2016design}. Resonant diffraction intensity was collected by a custom-designed CCD detector of $2148\,\times\,2052$ pixels (XCAM Co, ltd, UK).
Since small-angle scattering intensity is distributed near the transmitted direct beam, a tungsten beamstop was introduced to protect the detector for RSXS experiment, while in case of HERALDO the smaller aperture size allowed us to measure the holograms without using any beamstop.
% The longitudinal coherence of the monochromatic linearly polarized direct beam of 10\,$\mu$m more than two times exceeded the diameter of the sample aperture thus satisfying the oversampling condition \cite{miao2000possible}.
The magnetic field was applied parallel to the incident X-ray beam and perpendicular to the thin plate.  A He-flow-type cryostat was used to control the sample temperature in range from $\sim 15$\,K to $\sim 320$\,K, as measured by the Si diode thermometers attached next to the sample holder and cryostat head \cite{yamasaki2013diffractometer}. A radiation shield was wrapped around the sample holder to reduce the heating of the specimen from the warm environment.

\section{Results and discussions}

\begin{figure}
\includegraphics[width=8.5cm]{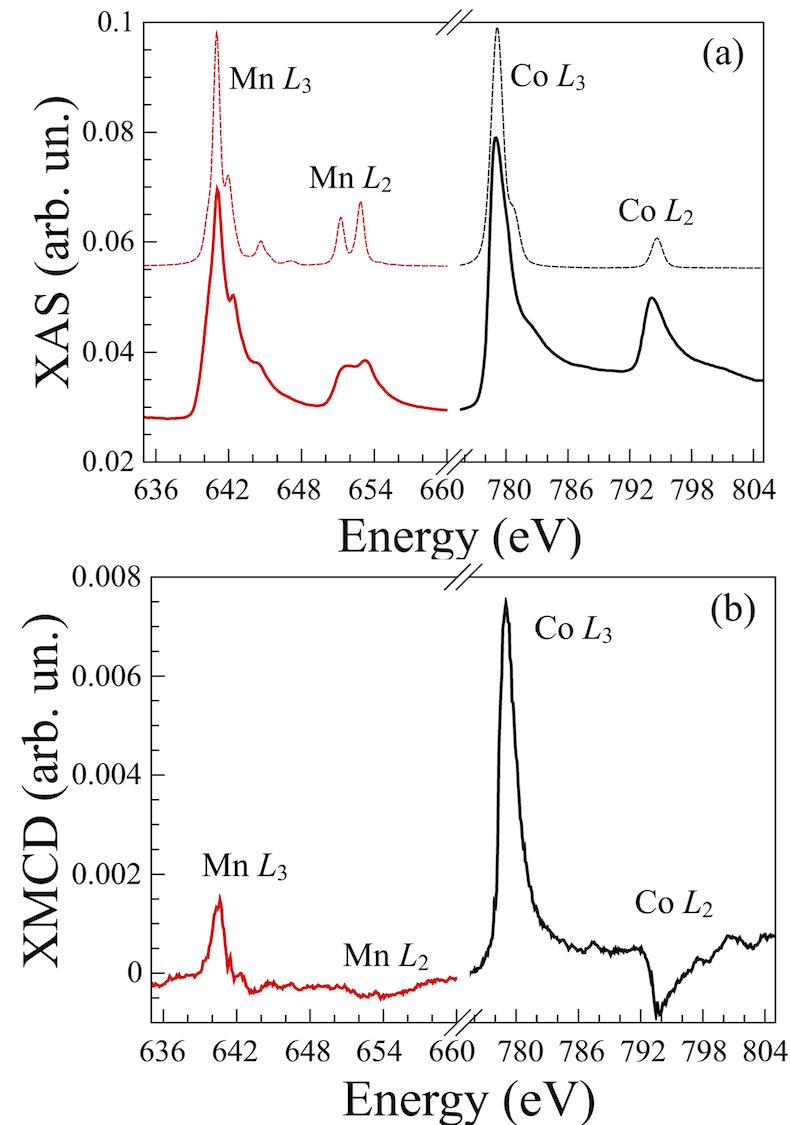}
\caption{(Color online) (a) XAS spectra of the bulk Co$_8$Zn$_8$Mn$_4$ near Co and Mn $L_{2,3}$ absorption edges: measured points are shown as the solid lines; calculated spectra are shown as the dashed lines. (b) XMCD signal. The measurements were performed at temperature $T=130$\,K at applied magnetic field $B=0.5$\,T.}
\label{Fig2}
\end{figure}

\begin{figure*}
\includegraphics[width=17cm]{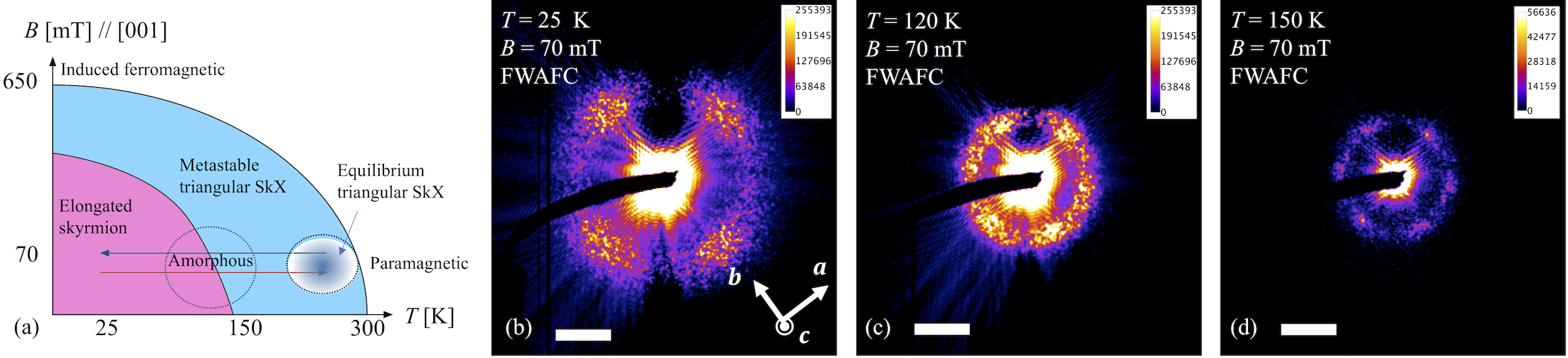}
\caption{(Color online) (a) Schematic phase diagram of Co$_8$Zn$_8$Mn$_4$ and the procedure of the resonant soft x-ray scattering (RSXS) measurements. The sample is field-cooled (FC) in the applied magnetic field $B = 70$\,mT from room temperature down to 25\,K (indicated by blue arrow) and then warmed back to 300\,K in the same field (indicated by red arrow). Coherent RSXS speckle patterns measured for Co$_8$Zn$_8$Mn$_4$ sample at $E=779$\,eV corresponding to $L_3$ absorption edge of Co at different temperatures (b) $T=25$\,K, (c) $T=120$\,K, (d) $T=150$\,K and applied field $B=70$\,mT. White scale bar corresponds to 0.05\,nm$^{-1}$.}
\label{Fig3}
\end{figure*}

The XAS signals averaged between RCP and LCP spectra near the $L_{2,3}$ edges of Mn and Co at $T=130$\,K are shown in Fig. \ref{Fig2}a. Surprisingly, despite the metallic nature of the Co$_8$Zn$_8$Mn$_4$ alloy, the Mn absorption shows a multiplet structure at the $L_3$ and $L_2$ edges. Well-resolved peaks at 1.3 and 3.5\,eV above the absorption maxima at $E=641.0$\,eV and a doublet structure at the $L_2$ edge, which splits into two maximums at $E=651.7$\,eV and $E=653.2$\,eV are clearly observable. Meanwhile, the Co $L_3$ and $L_2$ peak shapes at $E=780.0$\,eV and $E=794.1$\,eV are similar to the broad spectrum of metallic Co \cite{van1991strong,schwickert1998x,dhesi2001anisotropic}. This suggests that the fine structures of Mn $2p \rightarrow 3d$ transition should result from the localization of Mn $3d$ electrons rather than the oxidation of surface \cite{miyamoto2003soft,grabis2005element,gilbert2003multiple}: otherwise, the multiplet structure of Co $L_{2,3}$ edges should be either observed \cite{regan2001chemical,magnuson2002electronic}. To qualitatively illustrate the features of the measured XAS, simulations were performed using the small cluster approach in the Xclaim software tool \cite{fernandez2015xclaim}. Mn $3d^5$ was assumed as an initial state and Mn $2p^5$ $3d^6$ was the final configuration. The simulated XAS lines were broadened with a Lorentz function with the full width $\Gamma$ at half maximum (FWHM) of 1\,eV. The spectrum for Co was calculated using initial $3d^8$ and final $2p^5$ $3d^9$ configurations, which reproduces the measured data despite some broadening of the latter. The observed fine structure of the XAS spectrum of Mn also nicely corresponds to the one calculated from the multiplet effects (Fig. \ref{Fig2}a).

\begin{figure}
\includegraphics[width=8.5cm]{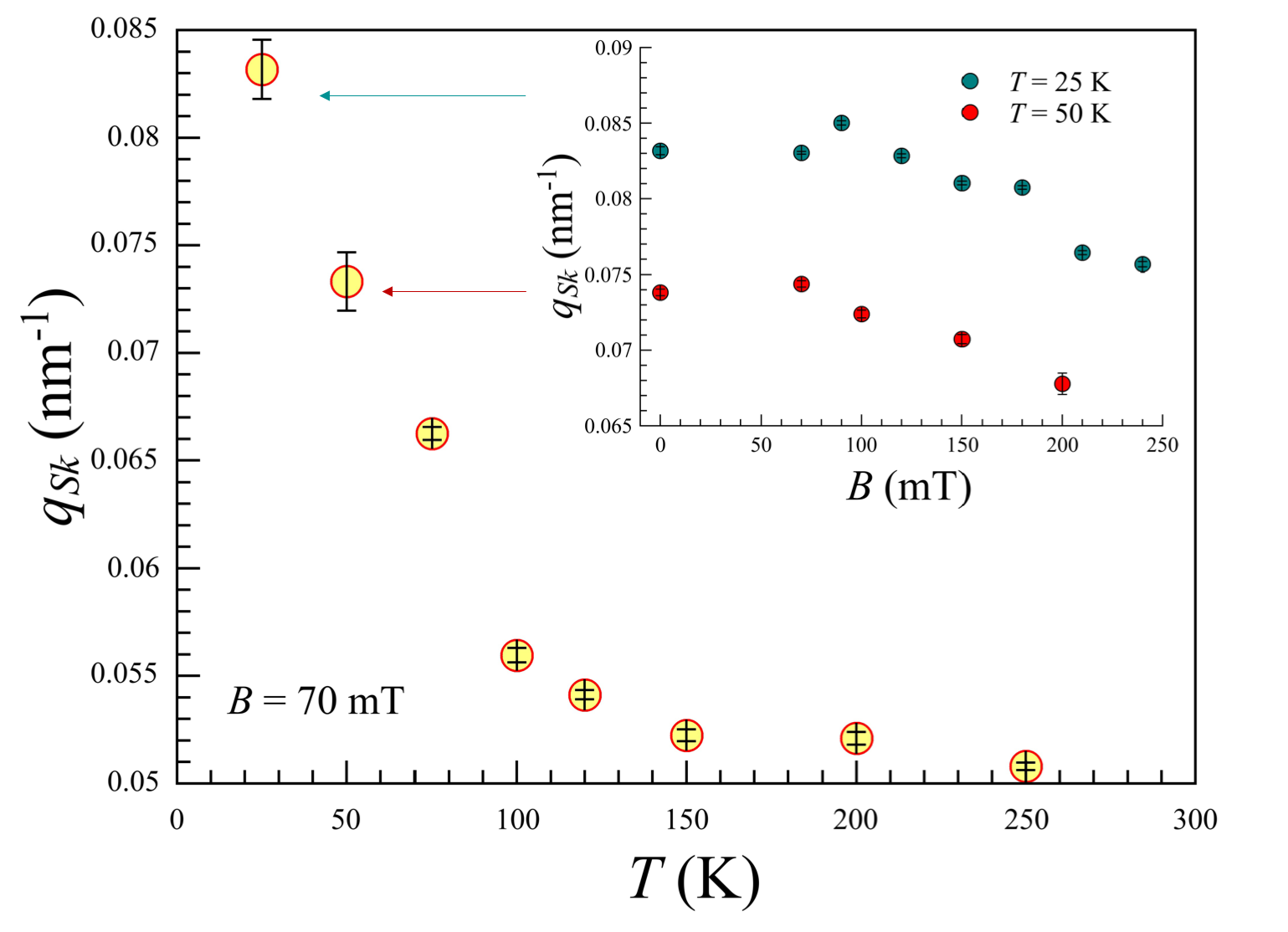}
\caption{(Color online) Temperature dependence of the modulation vector of skyrmion lattice $q_{Sk}$. Inset shows magnetic field dependence of $q_{Sk}$ measured at $T=25$\,K and $T=50$\,K.}
\label{Fig4}
\end{figure}

XMCD signal measured in a magnetic field of $B=0.5$\,T well above the saturation is shown in Fig. \ref{Fig2}b for both elements. Despite the magnitude of the XMCD measured at $L_{2,3}$ edges of Mn is about \sfrac{1}{5} times smaller than XMCD measured at Co $L_{2,3}$ edges, it is clear that the signs of the dichroic signals are the same. XMCD signal at Mn $L_2$ edge is notably suppressed, indicating quenching of the orbital moment.
The sum rule analysis \cite{o1994orbital} allows to estimate the orbital to spin moment ratio for Mn and Co as $\mu_l$(Mn)/$\mu_s$(Mn)=0.03 and $\mu_l$(Co)/$\mu_s$(Co)=0.0025.
Magnetic field dependence of the element-selective XMCD signals measured at 135\,K can be found in the Supplementary materials \cite{supplementary}. The signs and magnitudes of the XMCD signals indicate a ferromagnetic coupling of Co and Mn moments and a partial cancellation of Mn magnetization. This is in a good agreement with the magnetization measurements, which have shown the reduction of magnetization and critical temperature with increment of Mn concentration in $\beta$-Mn-type Co-Zn-Mn compounds \cite{tokunaga2015new}. One can assume that while the Co-Co and Co-Mn couplings are ferromagnetic, the Mn-Mn interaction should be antiferromagnetic. This scenario is also suggested by recent neutron diffraction study of the Co-Zn-Mn alloys in a wide composition range accompanied by the density functional theory (DFT) calculations \cite{bocarsly2019deciphering}.  This is in contrast to the case of Co-Mn alloys where Mn moments tend to align antiparralel to the host Co magnetization \cite{nakai1978magnetic,menshikov1985magnetic,wildes1992polarized}. On the other hand, the parent $\beta$-Mn compound shows strongly antiferromagnetic nearest-neighbor correlations in the $12d$ Mn sublattice, while strong ferromagnetic correlations between next-nearest-neighbors were found \cite{paddison2013emergent}.  Further, recent DFT calculations of the $\beta-$Mn-type Co-Zn-Mn suggested larger localization of the Mn atoms at $12d$ site than at $8c$, being consistent with our observations \cite{bocarsly2019deciphering}. For further quantitative discussions on the XAS and XMCD features, such as origin of multiplet spectrum of Mn, and determination of spin and orbital contributions, additional measurements of different Co-Zn-Mn concentrations and spectra calculations from \textit{ab initio} theory are highly desired.

Complex scattering factor $f$ for resonant magnetic X-ray scattering can be described as

\begin{equation}
f=(\bm{s}\cdot \bm{s}') f_c + i (\bm{s}\times \bm{s}')\cdot \bm{M}f_{m}^{1} + (\bm{s}\cdot\bm{M})(\bm{s}'\cdot\bm{M})f_{m}^{2},
\label{eq1}
\end{equation}
where $\bm{s}$ and $\bm{s}'$ are polarizations of the incident and scattered X-rays, respectively; $f_c$ is the charge scattering factor; $\bm{M}$ is the local magnetization; $f_{m}^1$ and $f_{m}^2$ are factors attributed to the magnetic scattering maximized at the resonant condition. The last term in Eq. (\ref{eq1}) containing scattering factor $f_{m}^2$ is quadratic in $\bm{M}$ and generally smaller than the other two terms \cite{hannon1988x}. Therefore the scattering patterns $(|f|^2)$ measured at resonant conditions mainly consist of the squares of charge and magnetic scattering factors $|f_c|^2$ and $|f_{m}^1|^2$, and their interference ($|f_c^* f_{m}^1|$). For linear polarization the measured intensities are dominated by the pure charge and magnetic scattering terms, while those measured using circular polarization contain charge-magnetic cross term \cite{hill1996resonant,eisebitt2003polarization}. Consequently, the intensity of magnetic scattering can be simply distinguished from the charge scattering by subtracting the diffraction pattern measured 1) at off-resonant condition and appropriately normalized or 2) in the field-induced polarized state of the sample or 3) above the critical temperature.

RSXS experiments were carried out at Mn ($E=640.5$\,eV) and Co ($E=779$\,eV) $L_{3}$ edges, where the magnetic scattering intensity was maximized. Far-field RSXS patterns were acquired with an exposition time of 1\,second (excluding readout time $\sim 0.4$\,s) and total 200\,expositions were averaged. In case of the off-resonant conditions, the absorption was reduced and the measurement time was consequently reduced by a factor of two. RSXS was measured in the temperature range between $T=25$\,K and $T=270$\,K during a field 
-warming-after-field-cooling (FWAFC) in $B=70$\,mT. The sample position was realigned after each temperature change to compensate the effect of the thermal contraction (expansion) of the sample manipulator with a buffer time between the measurements (in average $\approx 40$\,minutes) for stabilization. In the present experiment the linear polarization was used to minimize the influence of the charge-magnetic scattering interference term. In order to isolate the charge scattering contribution we have measured the scattering intensity at off-resonance ($E=645$\,eV and $E=785$\,eV) conditions.  
\begin{figure*}
\begin{center}
\includegraphics[width=15cm]{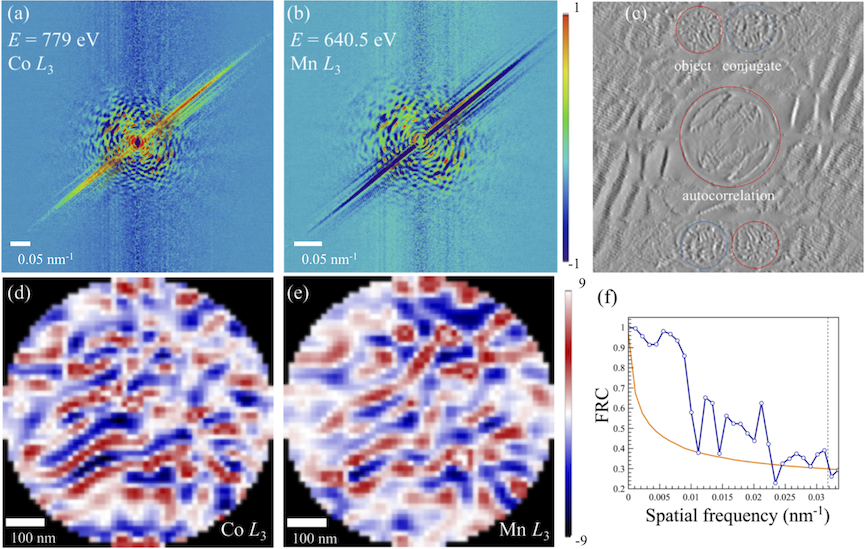}
\caption{(Color online) Differential scattering patterns  between LCP and RCP X-rays of energies (a) $E=779$\,eV and (b) $E=640.5$\,eV the taken at $T=20$\,K. The colorbar is given for both patterns in arbitrary units. Note that the strong linear patterns inclined upward right should be due to the incomplete subtraction of the charge scattering from the reference slit. (c) Fourier transform of (a) after applying linear differential filter and rotation of the image by 39$^\circ$. (d), (e) Magnetic texture recorded at Co and Mn $L_3$ edges, respectively. The colorbar is given for both images in arbitrary units. The region outside the field of view is filled with the black background manually for clarity. (f) Fourier ring correlation analysis of the reconstructed images for Co and Mn. Orange line represents calculated half-bit threshold and the dashed line indicates irreversible cross-point between the FRC function and half-bit threshold.}
\label{Fig5}
\end{center}
\end{figure*}

Typical coherent RSXS speckle patterns measured at $E=779$\,eV at temperatures $T=25$\,K, $T=120$\,K, and $T=150$\,K are shown in Fig.\ref{Fig3}. The missing area in the left part of each panel is due to the beamstop shadow. Tuning the energy to the Mn $L_3$ edge results in the scaling of the scattering pattern due to the difference in the photons wavelength. Magnetic scattering intensity is weaker at Mn edge, which can be explained by the lower Mn concentration in this compound, stronger absorption, and magnetization reduction, as expected from XMCD experiment. At room temperature and $T=150$\,K the small-angle scattering patterns demonstrate six-fold symmetry indicating single-domain triangular skyrmion lattice. As it has been already shown by SANS and LTEM experiments, the skyrmion lattice phase can be supercooled by a field cooling process down to the low temperatures \cite{karube2016robust,nakajima2017skyrmion,morikawa2017deformation}. According to the previous LTEM observations on a Co$_8$Zn$_8$Mn$_4$ thin plate, hexagonal skyrmion lattice to amorphous state transition is reversible and accompanied by the elongation of individual skyrmions along one of the principal crystallographic axes, while the skyrmion density is conserved \cite{morikawa2017deformation}.

Upon the field cooling (or FWAFC), the RSXS transforms to a homogeneous ring-like pattern corresponding to the intermediate "amorphous" phase at the temperature $T\approx100$\,K. Below $T\approx100$\,K four wide peaks appear at the coherent diffraction speckle pattern. Moreover, below the transition temperature, the skyrmion lattice parameter gradually decreases from $a_{Sk}=112$\,nm ($T=100$\,K) to $a_{Sk}=76$\,nm ($T=25$\,K). Temperature dependence of the magnetic modulation $q_{Sk}$ is shown in Fig. \ref{Fig4}. $q_{Sk}$-vector is also dependent on magnetic field similar to the bulk case \cite{karube2016robust} (inset in Fig. \ref{Fig4}). Elongation of the $q_{Sk}$ vector is, presumably, caused by increasing antiferromagnetic interaction between Mn ions superimposed on the helical order. Upon cooling from $T=100$\,K to $T=25$\,K a coherent RSXS speckle pattern with four-fold symmetry can be observed. According to the previous results, the hexagonal skyrmion lattice \cite{karube2016robust} is recovered by applying stronger magnetic field of 300-500\,mT at low temperature. Indeed, the magnitude of the $q_{Sk}$ vector tends to shrink upon increment of the magnetic field (inset in Fig. \ref{Fig4}). However, in present experiment the magnetic scattering intensity arising from isolated skyrmions is hardly distinguishable in presence of the background even when the charge scattering is subtracted. Corresponding RSXS patterns can be found in the Supplementary materials \cite{supplementary}.

The integrated speckle pattern intensities are demonstrating similar features for Co and Mn except the difference in signal-to-noise ratio, which is better at $E=779$\,eV. Radially integrated azimuthal profiles of the scattering patterns measured using soft X-rays with energies $E=779$\,eV and $E=640.5$\,eV can be found in the Supplementary materials \cite{supplementary}. 

Fourier transform holography with extended reference allows to reconstruct the real-space image by single Fourier transform of the measured pattern multiplied by the linear differential operator in direction parallel to the reference slit. Circularly polarized soft X-rays with energies matching to Co and Mn $L_3$ edges were used for HERALDO experiment. Far-field diffraction patterns were collected with RCP and LCP to produce the interference pattern and reconstruct the magnetic texture. In this case information about the magnetic texture was simply encoded in the interference term $f_c^* f_{m}^1$ between charge scattering arising from the reference slit and magnetic scattering from the Co$_8$Zn$_8$Mn$_4$ sample. The difference between the holograms taken with the two opposite X-ray helicities provided the isolated interference term which could be inverted to the real-space image via single Fourier transform and linear differential operator \cite{guizar2007holography}. HERALDO patterns were acquired with an exposition time of 6\,s (excluding readout time $\sim 0.4$\,s), and total 200 expositions were averaged, resulting in a total acquisition time of 20 minutes for each hologram. The sample was cooled down to the temperature $T=20$\,K while the magnetic field of $B=70$\,mT was applied along $[001]$ direction parallel to incident soft X-ray beam propagation vector.

\begin{figure*}
\includegraphics[width=16cm]{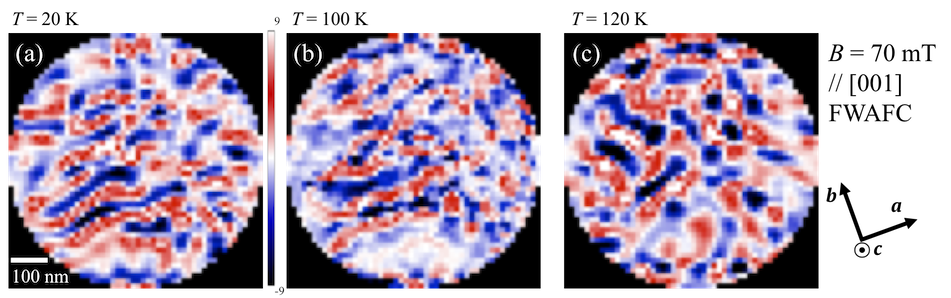}
\caption{(Color online) Real-space magnetic textures imaged at Co $L_3$ edge ($E=779$\,eV) at temperatures (a) $T=20$\,K, (b) $T=100$\,K and (c) $T=120$\,K and applied magnetic field $B=70$\,mT.}
\label{Fig6}
\end{figure*}

Differences between the diffraction patterns taken with the two opposite X-ray polarizations at photon energies $E=779$\,eV and $E=640.5$\,eV are shown in Figs. \ref{Fig5}a and b, respectively. The highest harmonics of the interference pattern can be found at $q_{max}=0.2$~nm$^{-1}$ which corresponds to the real-space resolution of 32\,nm. The difference between Fourier transform images taken with RCP and LCP at Co $L_3$ edge after applying the differential filter in the slit direction is shown in Fig. \ref{Fig5}c. Reconstructed real-space image shows the sample autocorrelation and two object-reference cross-correlations delivered by both ends of the slits, as well as their complex conjugates. Magnetic contrast is inverted between the reconstructed object and conjugate. By taking into account the widths of magnetic domains where the local magnetization points antiparallel (parallel) to the applied field $B=70$\,mT at different temperatures (Fig. \ref{Fig6}), we conclude that the regions with the negative (positive) values in Figs. \ref{Fig5}d and e correspond to magnetization pointing antiparallel (parallel) to the field.

Figures \ref{Fig5}d and e show magnification of the sample-reference cross-correlations taken from reconstructions of holograms taken for Co and Mn, respectively. Due to the difference of the wavelengths, the real-space image corresponding to magnetic texture of Mn ions was scaled by factor $E_1/E_2=779/640.5\approx1.22$. The magnetic structures exhibited by magnetic elements Mn and Co are similar to each other within the resolution limit. Signal-to-noise ratio is worse in case of the measurement at Mn $L_3$ edge due to the higher absorption of soft X-rays at $E=640.5$\,eV and smaller concentration of Mn atoms compared to Co. Different magnitude of the magnetic moment between Co and Mn is also a reason for this. Therefore, the real-space image reconstructed for Mn atoms is slightly blurred (Fig. \ref{Fig5}e). Additionally the reconstruction resolution is estimated from the Fourier ring correlation (FRC) \cite{banterle2013fourier}. By using FRC the spatial frequency dependence of the cross-correlation of two real-space magnetic textures obtained from the reconstructions of Co and Mn HERALDO patterns (Fig. \ref{Fig5}f) is analyzed. The resolution was calculated at the point where the FRC curves irreversibly cross the half-bit threshold \cite{van2005fourier}. The real-space resolution that was found according to this criteria is similar to the resolution of 32\,nm determined from the highest interference harmonics observed at the Fourier-space pattern.

\begin{figure*}
\includegraphics[width=17cm]{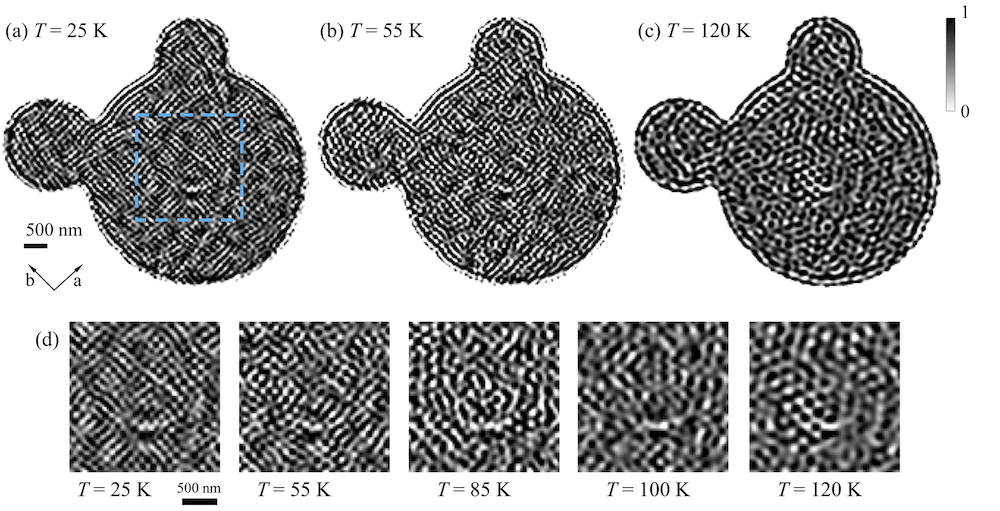}
\caption{(Color online) Reconstruction of coherent RSXS patterns at (a) $T=25$\,K, (b) $T=55$\,K and (c) $T=120$\,K collected at Co L$_3$ edge. The colorbar for intensity ($z$-scale) is given in arbitrary units. (d) Magnification of the region highlighted by the blue square in (a) is shown for the the different temperatures.}
\label{Fig7}
\end{figure*}

Elongated skyrmions can be observed at the holograms taken both at $E=779$\,eV and $E=640.5$\,eV indicating similar magnetic texture of Co and Mn sub-lattices, which coincides with the XMCD and RSXS data. Magnetic holograms were recorded during a FWAFC procedure up to 120\,K at which the magnetic scattering is highly reduced due to the thermal shrinkage of the magnetic moments. Corresponding evolution of the magnetic texture measured at Co absorption edge is shown in Fig. \ref{Fig6}. The transformation of the elongated skyrmions to more conventional shape with corresponding expansion takes place at $T=120$\,K (Fig. \ref{Fig6}c). Unfortunately, due to a thermal shrinkage of magnetic moments the real-space reconstruction of magnetic texture at higher temperatures can be hardly distinguished from the background fluctuations.

\begin{figure*}
\includegraphics[width=17cm]{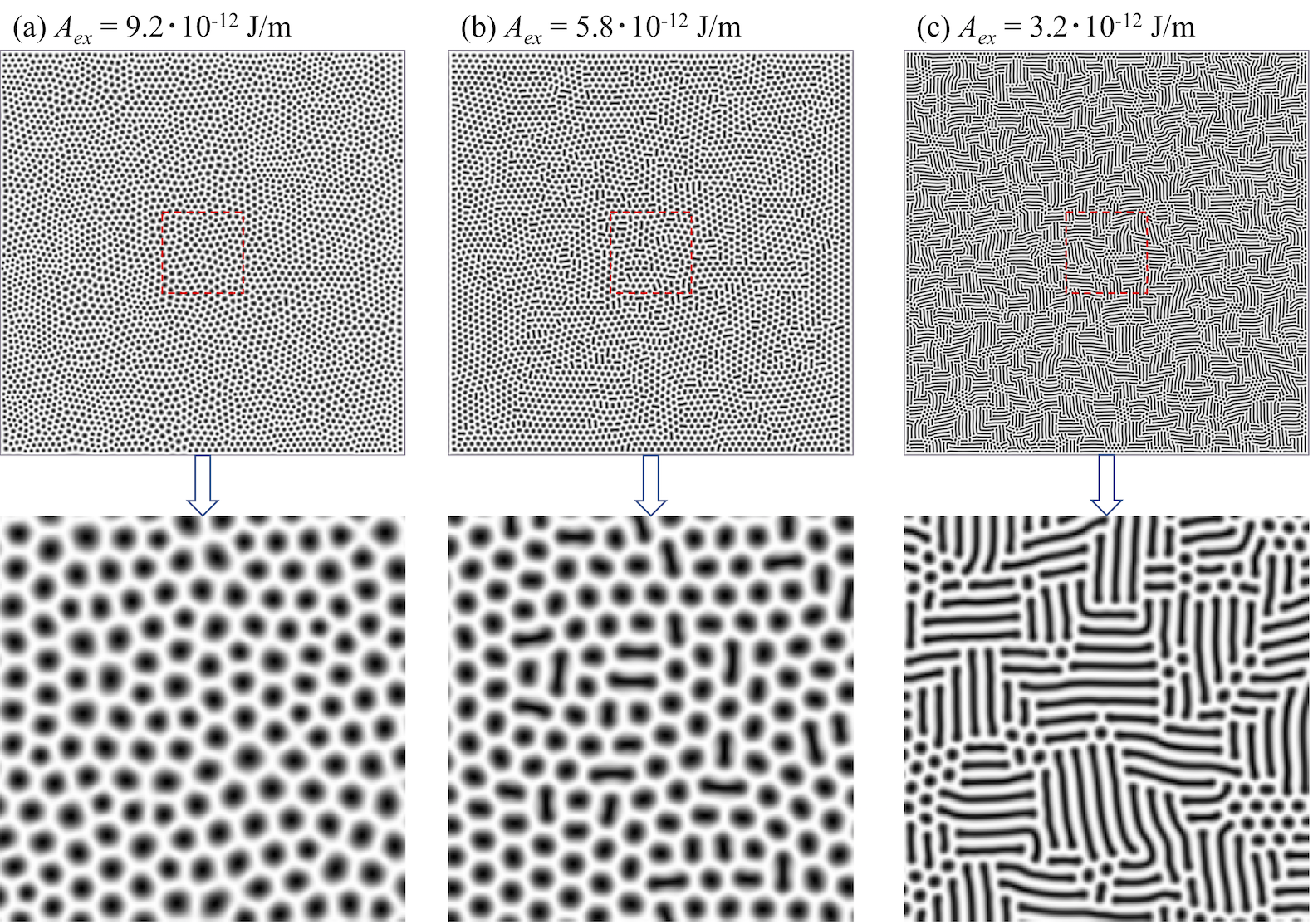}
\caption{(Color online) Micromagnetic simulations of Co$_8$Zn$_8$Mn$_4$ $10\times10$\,$\mu$m$^2$ thin plate: $z-$projection of the magnetic texture is shown for various exchange from stiffness parameters $A_{ex}=9.2\cdot10^{-12}$\,J/m (a), $5.8\cdot10^{-12}$\,J/m (b), $3.2\cdot10^{-12}$\,J/m (c). Bottom panels show magnification of the rectangular regions highlighted by the corresponding red boxes.}
\label{Fig8}
\end{figure*}

Since the sample area probed by HERALDO was limited by the aperture size, we additionally performed a real-space inversion off the measured RSXS patterns by the iterative phase retrieval. Square root of the measured magnetic scattering intensity $\sqrt{I_{m}}$ shown in Fig. \ref{Fig3} isolated from the charge scattering as described in Ref. \cite{ukleev2018coherent}, was used as real part of the Fourier-space constraint. Fixed sample aperture size, shape and orientation were used as real-space support. Pixels, that were missing at the center of diffraction pattern due to the beamstop shadow and subtracted charge scattering, were substituted by the Fourier transform of the support. This approach is similar to the substitution used in the guided hybrid input-output (HIO) algorithm \cite{chen2007application}. We used the implementation of HIO algorithm \cite{fienup1982phase} with a feedback parameter $\beta=0.9$ with assuming positivity and reality constraints for the real-space pattern. Phase-retrieval algorithm tended to stagnate to the local minima after 500 iterations. For each coherent scattering pattern the final real-space image was calculated from average of 500 algorithm trials with random initial phases. Reliability of the reconstructed real-space images was qualitatively and quantitatively examined by comparing the reconstructions performed for the data measured at Co and Mn $L_3$ edges, respectively. Indeed, based on the results of the XMCD and HERALDO experiments the same orientation of the local magnetic moments of Co and Mn atoms was expected.

\begin{figure}
\includegraphics[width=8.5cm]{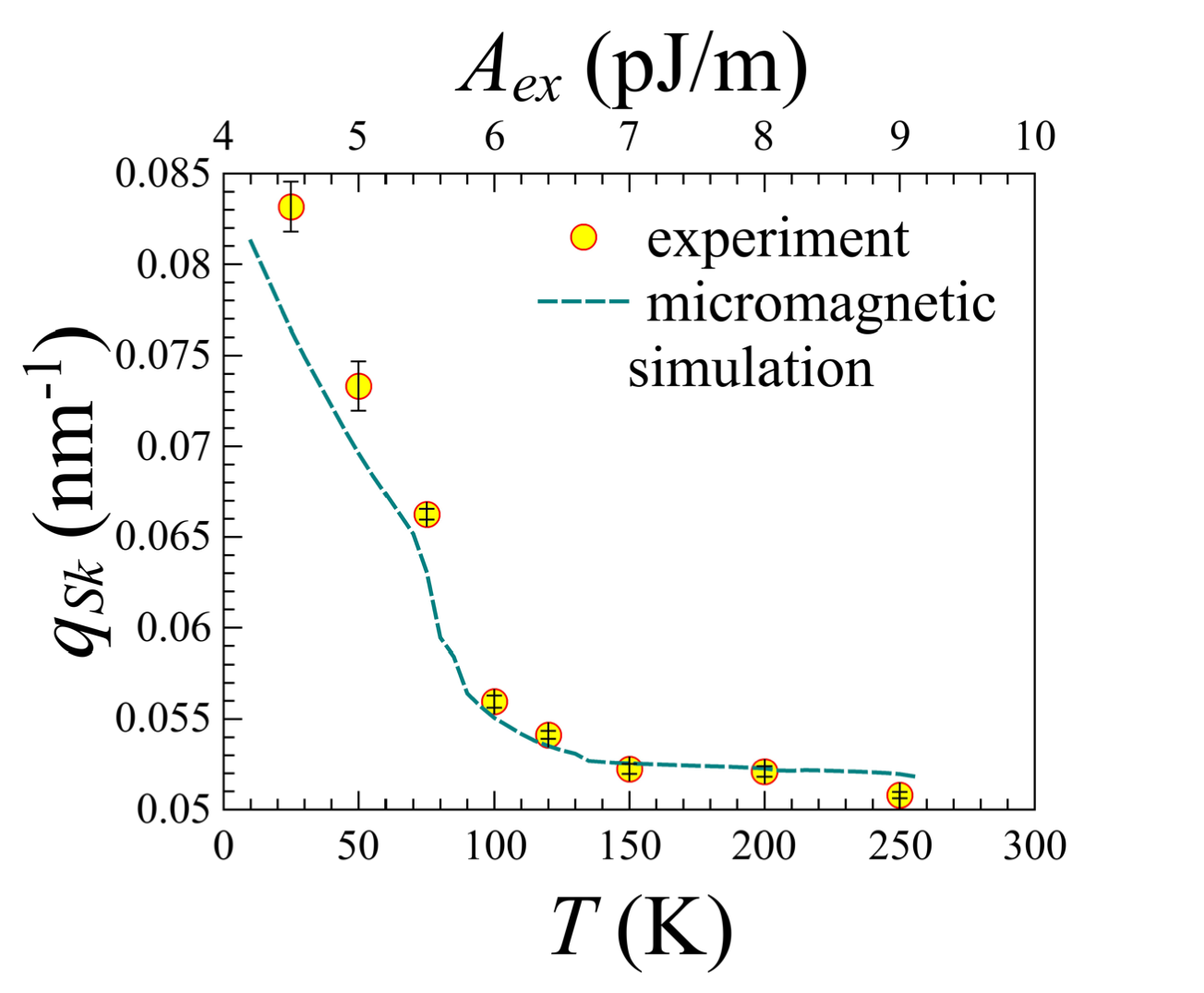}
\caption{(Color online) Measured dependence of $q_{Sk}$ over temperature (symbols) and calculated dependence of $q_{Sk}$ over the exchange stiffness ($A_{ex}$) parameter.}
\label{Fig9}
\end{figure}

\begin{figure*}
\includegraphics[width=17cm]{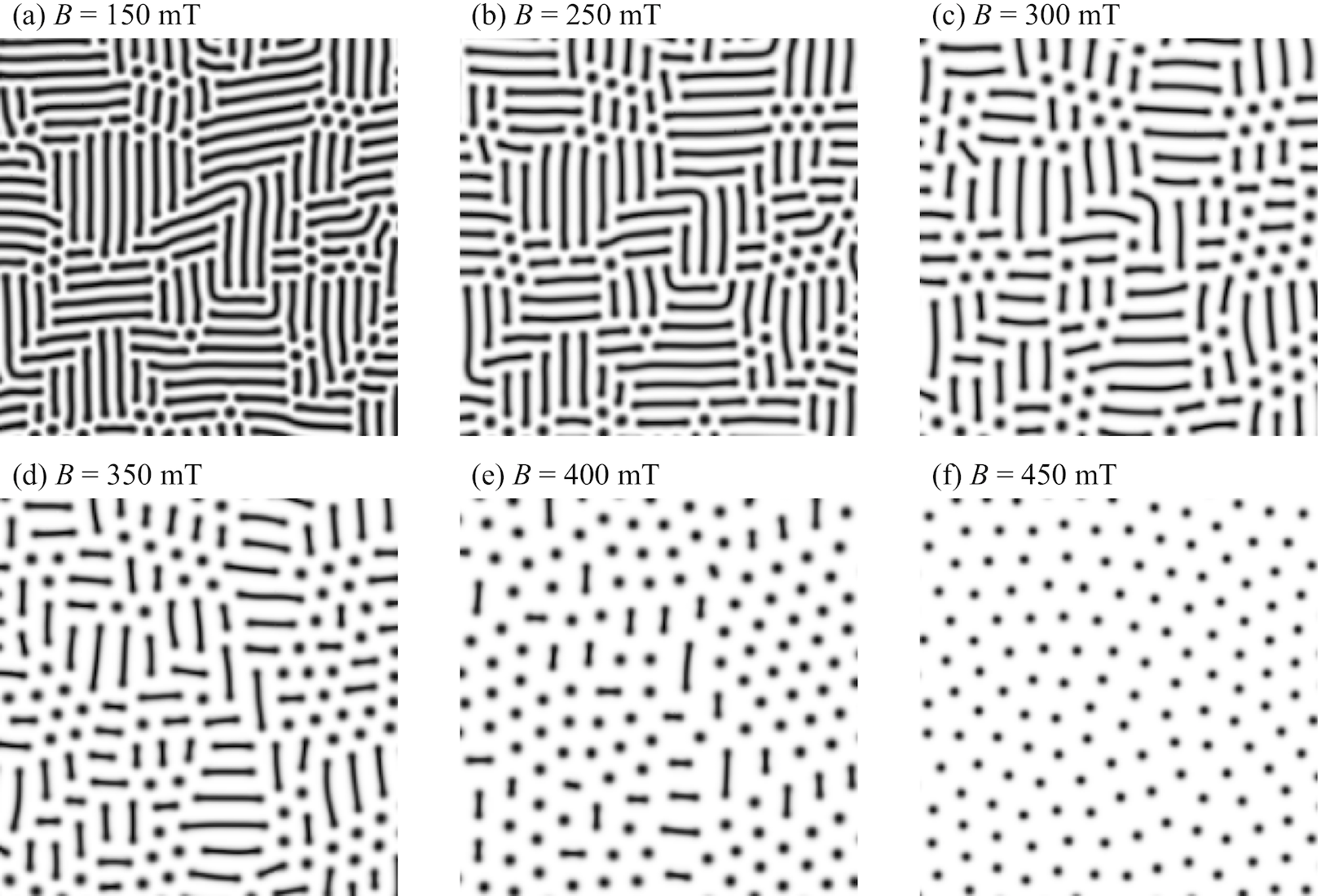}
\caption{(Color online) Magnetic field evolution of the $z-$projection of magnetization derived from the micromagnetic simulation of Co$_8$Zn$_8$Mn$_4$ thin plate in magnetic field $B=150$\,mT (a), $B=250$\,mT (b), $B=300$\,mT (c), $B=350$\,mT (d), $B=400$\,mT (e), $B=450$\,mT (f) showing the hexagonal SkX restoration.}
\label{Fig10}
\end{figure*}

To quantitatively estimate the reliability of the reconstructions performed for Co and Mn we have introduced the two-dimensional function 
$$R(i,j)=1-\frac{|\psi_{E_1}^{*(i,j)} - \psi_{E_2}^{*(i,j)}|} {|\psi_{E_1}^{*(i,j)}|},$$
where $\psi_{E_1}^{*(i,j)}$ and $\psi_{E_2}^{*(i,j)}$ are values of $(i,j)$ pixels of the reconstructed real-space patterns normalized by maximum value for corresponding image at Co($E_1$) and Mn($E_2$) $L_3$ edges. This normalization is needed to compensate the difference in total magnetic contrast for Mn and Co patterns and transmission coefficient for the photons with different energies. Figure \ref{Fig7} shows the real-space magnetic patterns reconstructed from data taken at Co absorption edge at different temperatures. At $T=25$\,K and $T=55$\,K (Figs. \ref{Fig7}a,b) an array of magnetic skyrmions elongated along one of the crystallographic axes can be observed. As the temperature increases, some of the elongated skyrmions transform to the conventional circular-shaped (Fig. \ref{Fig7}c), denoted as amorphous state. Unfortunately, similarly to the HERALDO experiment, the magnetic scattering intensity gradually decreases upon warming due to the thermal shrinkage of the magnetic moments. As a result, the signal to noise ratio is insufficient to reliably reconstruct the real-space image of the single-domain triangular skyrmion lattice at $T>120$\,K. In the Supplementary information \cite{supplementary} we show a few examples of the reconstructions performed for the skyrmion lattice and multidomain helical phases at higher temperatures. However, the real-space images obtained for these conditions correspond to the local minima in the solution space and vary sufficiently for the different starting phases. Two-dimensional reliability maps $R(i,j)$ for successful reconstructions can be found in the Supplementary information \cite{supplementary}. Here we just note that the averaged reliability value $R=\frac{1}{N}\sum R(i,j)$, where $N$ is the sample area in pixels, is not less than 89\% for the datasets measured at 25\,K.

From this aspect, the results of lensless imaging by iterative phase retrieval and HERALDO are consistent with each other. Both techniques provide a spatial resolution of 30\,nm and sensitive to the signal-to-noise ratio. Data analysis in case of coherent diffraction imaging is more sophisticated compared to HERALDO, but the latter requires advanced sample fabrication. Further attempts to improve the nanofabrication routine are required to obtain better real-space resolution. 

The experimental results for the temperature- and field-dependent transformation of the skyrmion lattice obtained by soft X-ray scattering and imaging agree well with the results of Landau-Lifshitz-Gilbert (LLG) simulations calculated with realistic parameters for Co$_8$Zn$_8$Mn$_4$ thin plate by using mumax$^3$ package \cite{vansteenkiste2014design}. As hinted by present XMCD experiment and previous magnetization measurements \cite{tokunaga2015new}, the effective ferromagnetic exchange interaction in Co$_8$Zn$_8$Mn$_4$ system may decrease with temperature due to antiferromagnetic correlations of Mn sub-lattice. Moreover, manifestation of the cubic anisotropy is considerable at low temperatures. For the simulation we used exchange stiffness $A_{ex} = 9.2$\,pJ/m and DMI constant $D=0.00053$\,J/m measured experimentally by microwave spin-wave spectroscopy \cite{takagi2017spin}, and cubic anisotropy in the form of an effective field \cite{vansteenkiste2014design}:
\begin{equation*}
\begin{aligned}
\mathbf{B}_{anis} &= - 2 K_c / M_{s} \big(((\mathbf{c}_{1} \cdot \mathbf{m})\mathbf{c}_{1})((\mathbf{c}_{2} \cdot \mathbf{m})^2 + (\mathbf{c}_{3} \cdot \mathbf{m})^2)+ \\
&((\mathbf{c}_{2} \cdot \mathbf{m})\mathbf{c}_{2})((\mathbf{c}_{1} \cdot \mathbf{m})^2 + (\mathbf{c}_{3} \cdot \mathbf{m})^2)+ \\
&((\mathbf{c}_{3} \cdot \mathbf{m})\mathbf{c}_{3})((\mathbf{c}_{1} \cdot \mathbf{m})^2 + (\mathbf{c}_{2} \cdot \mathbf{m})^2)\big),
\end{aligned}
\end{equation*}
where $K_c = 5000$\,J/m$^3$ is the 1$^{st}$-order cubic anisotropy constant determined from electron spin resonance (ESR) experiment \cite{kezmarki2018}; $\mathbf{c}_{1}$, $\mathbf{c}_{2}$ and $\mathbf{c}_{3}$ is a set of mutually perpendicular unit vectors indicating the anisotropy directions (cubic axes), and $M_s=350$\,kA/m is the saturation magnetization \cite{takagi2017spin}. A two-dimensional 200\,nm-thick plate with area of $10\times10$\,$\mu$m$^2$ with open boundary conditions and elementary cell size $5\times5\times200$\,nm$^3$ was simulated. To follow the experimental field-cooling protocol, we started the simulation with random initial spin configuration in the applied field along $[001]$ direction. The field needed for robust SkX formation was set to $B=150$\,mT. The discrepancy between the observed and calculated field values is, presumably, caused by demagnetization effects. The characteristic helical pitch $\lambda=4\pi A_{ex}/D$ and skyrmion size are determined by the ratio of the exchange stiffness $A_{ex}$ and the Dzyaloshinskii constant $D$ \cite{bak1980theory}. Therefore it is reasonable to assume two possible scenarios: 1) the temperature-dependent $q$-vector variation is caused by changing exchange integral due to the antiferromagnetic correlations in Mn sub-lattice towards lower temperatures; 2) the emergence of the $q$-vector elongation can be also caused by enhancement of antisymmetric DMI due to the local non-complanar structure of frustrated Mn. At the moment, the experimental data does not allow to unambiguously distinguish between variations of $A_{ex}$ and $D$. To mimic the first scenario, we reduced the exchange stiffness parameter $A_{ex}$ from the measured value $9.2$\,pJ/m to $3.2$\,pJ/m in a linear fashion, while the magnetic field $B=150$\,mT remained constant. Relaxation time of 10 ns was introduced between each step for magnetic texture stabilization. Upon gradual decrease of the exchange interaction in the system, the skyrmion lattice exhibits deformation similar to the LTEM and soft X-ray experiments (Figs. \ref{Fig7}a--c). Elongation of the skyrmions is manifested due to the overall reduction of the exchange interaction, while its directionality along the cubic axes is dictated by the cubic anisotropy. Notably, this deformation takes place via directional expansion of each topological vortex, but not by merging of the neighboring skyrmions -- in the latter case the total topological charge of the system would decrease, which is opposite to the topological charge conservation scenario reported earlier \cite{morikawa2017deformation}. Radial averages of the two-dimensional fast Fourier transformation (FFT) patterns of out-of-plane projection of magnetization distribution was used to calculate the $q$-vector magnitude dependence for the resultant magnetic textures. The simulated variation in $q_{Sk}$ induced by change in the exchange stiffness parameter $A_{ex}$ is in the very good qualitative agreement with the temperature dependence (Fig. \ref{Fig9}). The shoulder in Fig. \ref{Fig9} at $A_{ex}$ between 5 -- 6\,pJ/m corresponds to the smooth transition from the disk-shaped to elongated skyrmions. Therefore we assume that the linear decrease of the exchange interaction in Co$_8$Zn$_8$Mn$_4$ takes place with decreasing temperature. In general, this consideration is consistent with the previous studies of the $\beta-$Mn alloys, that have shown presence of antiferromagnetic correlation of moments in the $12d$ Mn sublattice \cite{shiga1994polarized,nakamura1997strong,stewart2009magnetic,paddison2013emergent,bocarsly2019deciphering}. We plan to directly address to this question in our future studies. For example, recently developed spin-wave sensitive SANS technique allows to directly probe $A_{ex}$ of a bulk specimen as a function of temperature \cite{grigoriev2015spin}.

LLG simulation was either used to probe the magnetic field evolution of the elongated skyrmion texture. The "field-cooled" magnetic texture shown in Fig. \ref{Fig10}c with $A_{ex}=3.2$\,pJ/m  was used. Except the discrepancy between the field  magnitude values, the result is consistent with present X-ray and previous LTEM experiments -- restoration of the hexagonal lattice of circular skyrmions from the deformed state by ramping the magnetic field was successfully reproduced (Figs. \ref{Fig10}a--f).

\section{Conclusion}

In conclusion, by means of element-selective soft X-ray circular magnetic dichroism we have revealed the ferromagnetic arrangement of Co and Mn ions in a room-temperature skyrmion-hosting compound Co$_8$Zn$_8$Mn$_4$. Moreover, by using the coherent resonant small-angle soft X-ray scattering and holography with extended reference we can conclude that the topological magnetic texture is the same for both type of atoms in whole temperature range above $T_g$ that is reliable for real-space reconstruction. Our results are consistent with each other and with the previous neutron scattering and Lorentz microscopy experiments and shows the transition from hexagonal skyrmion crystal to elongated skyrmion state that is accompanied by deformation of the individual vortices. Micromagnetic simulation suggests that such transition is driven by decreasing exchange interaction in the system and effect of the cubic anisotropy. This effective decrease of the ratio of symmetric exchange interaction to antisymmetric Dzyaloshinskii-Moriya mimics low-temperature antiferromagnetic frustration of Mn sub-lattice. At lower temperature, antiferromagnetic correlations of Mn atoms is superimposed onto the long-range helical (skyrmion) modulation, resulting in shortening of the helical pitch and deformation of skyrmions. However, this effect is reversible and hexagonal skyrmion lattice from elongated skyrmion state can be restored by increasing magnetic field even when the exchange stiffness is reduced, as learned from the micromagnetic simulation and previous experiments \cite{karube2016robust,morikawa2017deformation}.

We have demonstrated first to our knowledge lensless soft X-ray imaging of the magnetic texture at cryogenic temperatures and applied magnetic fields. HERALDO imaging was used with the circularly polarized soft X-rays, while coherent diffraction imaging was performed with a linearly polarized beam. Both methods did not require focusing X-ray optics to perform magnetic imaging with resolution of few tens of nanometers. Practically, the signal to noise ratio sufficient for the successful reconstruction was achieved only in the temperature range from $T=20$\,K to $T=120$\,K due to the overall decay of the intensity of the magnetic scattering and charge-magnetic interference with increasing temperature.

Soft X-ray imaging methods allow to simultaneously obtain element-selective real-space information and will be useful for further investigations of non-trivial magnetic textures in thin plates of polar magnets, since N\'eel-type skyrmions produce no contrast in Lorentz transmission electron microscopy.

\section*{Acknowledgments}
The authors wish to acknowledge P. Gargiani and BOREAS beamline staff for the technical assistance. We also thank T. Honda for providing the membranes. Soft X-ray scattering experiments were performed as a part of the proposals no.: 2015S2-007 (Photon Factory) and 2016081774 (ALBA Synchrotron Light Laboratory).
This research was supported in part by PRESTO Grant Number JPMJPR177A from Japan Science and Technology Agency (JST), "Materials research by Information Integration" Initiative (MI$^2$I) project of the Support Program for Starting Up Innovation Hub from JST, the Japan Society for the Promotion of Science through the Funding Program for World-Leading Innovative R\&D on Science and Technology (FIRST Program), and JSPS KAKENHI Grant Number 16H05990. V.U. acknowledges funding from the SNF Sinergia CDSII5-171003 NanoSkyrmionics. M. V. acknowledges additional funding to the MARES endstation by grants MICINN ICTS-2009-02, FIS2013-45469-C4-3-R and  FIS2016- 78591-C3-2-R (AEI/FEDER, UE).

\end{document}